\documentclass{article}
\usepackage{spconf,amsmath,graphicx}
\usepackage{multirow}
\usepackage[table,xcdraw]{xcolor}

\usepackage{hyperref}
\graphicspath{{.}{figs/}}
\DeclareGraphicsExtensions{.pdf, .png, .jpg}

\usepackage{multirow}
\usepackage{tabularx}
\usepackage{array}
\newcolumntype{Z}{>{\centering\let\newline\\\arraybackslash\hspace{0pt}}X}

\usepackage[font=normal,labelfont=bf]{caption}
\usepackage{enumitem}
\setlist[itemize]{align=parleft,left=0pt..1em}

\def\eb{{\mathbf e}}
\def\zerob{{\mathbf 0}}

\newcommand{\TODO}[1]{{\bf [ TODO: #1 ]}}

\newcommand{\OMIT}[1]{}

\title{Universal neural vocoding with Parallel WaveNet}
\name{Yunlong Jiao$^{\star}$, Adam Gabry\'{s}$^{\star}$, Georgi Tinchev$^{\dagger}$, Bartosz Putrycz$^{\star}$, Daniel Korzekwa$^{\star}$, Viacheslav Klimkov$^{\star}$\thanks{Corresponding author email: vklimkov@amazon.pl. Work done while Georgi Tinchev was an intern at Amazon. We would like to thank Alexis Moinet and Vatsal Aggarwal for insightful research discussions.}}
\address{$^{\star}$ Amazon.com \qquad $^{\dagger}$ University of Oxford}
\begin{document}
%\ninept
%
\maketitle
\begin{abstract}
We present a universal neural vocoder based on Parallel WaveNet, with an additional conditioning network called Audio Encoder.
Our universal vocoder offers real-time high-quality speech synthesis on a wide range of use cases.
We tested it on 43 internal speakers of diverse age and gender, speaking 20 languages in 17 unique styles, of which 7 voices and 5 styles were not exposed during training.
We show that the proposed universal vocoder significantly outperforms speaker-dependent vocoders overall.
We also show that the proposed vocoder outperforms several existing neural vocoder architectures in terms of naturalness and universality.
These findings are consistent when we further test on more than 300 open-source voices.%\footnote{Audio samples will be made publicly available upon a future version of the manuscript (copyright issues to be resolved).\OMIT{Audio samples are available at: \url{https://www.amazon.science/blog/}.}}
\end{abstract}
\begin{keywords}
Neural vocoder, Text-to-speech, Scalability
\end{keywords}
\vspace{-0.5cm}%
\section{Introduction}
\vspace{-0.3cm}%
\label{sec:intro}

As voice-based human-machine interaction becomes an increasingly crucial part in artificial intelligence, text-to-speech (TTS)\OMIT{, or synthesising natural sounding speech in real time,} remains an important yet challenging problem.
\OMIT{
	To model complex short-term and long-term dependencies of audio samples in speech waveforms, one has to operate at a temporal resolution of at least 16 kHz.
}
While some TTS systems synthesise speech from normalised text or phonemes in an end-to-end manner \cite{Ping2019ClariNet, Donahue2020End}, most TTS systems address the problem in a two-step approach.
The first step transforms text to a lower-resolution intermediate representations, such as time-aligned acoustic features \cite{Arik2017Deep}, or spectral features such as mel-spectrogram \cite{Wang2017Tacotron, Shen2018Natural}.
The second step transforms the intermediate representations to high-fidelity audio signal using a model, referred to as \textit{vocoder}.

\OMIT{
	Vocoding is a challenging task, since mel-spectrogram is a lossy representation of speech signals where critical information such as phase is absent.
	Traditional phase restoration algorithms such as Griffin-Lim \cite{Griffin1983Signal}, while capable of generating intelligible speech from their spectral representations, usually produce strong robotic artefacts due to the oversimplified statistical assumption on phase reconstruction.
}
State-of-the-art vocoders are neural network-based generative models \cite{Oord2016WaveNet, Kalchbrenner2018Efficient, Oord2018Parallel, Prenger2019Waveglow, Kumar2019MelGAN, Yamamoto2020Parallel}.
\OMIT{
	These neural vocoders can typically be categorised into two families: autoregressive and non-autoregressive.
	Autoregressive neural vocoders such as WaveNet \cite{Oord2016WaveNet} and WaveRNN \cite{Kalchbrenner2018Efficient} are often very high-quality vocoders with the caveat of inherently slow inference speed, since audio samples must be generated sequentially.
	Therefore, autoregressive neural vocoders are usually not suitable for real-time applications.
	Recently, non-autoregressive neural vocoders received a lot of attention, since their architectures are usually highly parallelisable, resulting in an inference speed typically orders of magnitude higher than autoregressive vocoders on modern deep learning hardware \cite{Oord2018Parallel, Prenger2019Waveglow, Kumar2019MelGAN, Yamamoto2020Parallel}.
	Like most machine learning technologies, neural vocoders are at a significant cost of computation and data requirement.
}
Neural vocoders are capable of synthesising natural-sounding speech, but typically prone to overfitting to the training data, and do not generalise well to unseen voices \cite{Arik2019Fast}.
Training speaker-dependent vocoders require significant computational resources and large amounts of audio data for each target speaker \cite{Wang2018Comparison}.
The need for a high-quality speaker-independent vocoder, or so-called \textit{universal} vocoder, is key to scaling up production of TTS systems that are specifically designed to support many voices.

A few recent studies investigated the possibility of building universal vocoders.
There were some early reports that speaker-independent vocoders underperform speaker-dependent vocoders \cite{Barbany2018Multi, Song2018Speaker}.
Deep Voice 2 \cite{Gibiansky2017Deep} modelled speaker identities by using trainable speaker embeddings in their vocoder as part of a multi-speaker TTS system.
These systems require to model speaker identities explicitly, hence cannot handle unseen speakers out-of-the-box.
\OMIT{
	Liu et al. \cite{Liu2018WaveNet} applied speaker adaptation methods by first training a multi-speaker vocoder, and then fine-tuned it using the limited data available per target speaker.
	However, it still requires to maintain a vocoder for each speaker during inference.
}
There are also reports of neural vocoders that are capable of synthesising unseen speakers or styles without having to explicitly model speaker identities \cite{Hayashi2017investigation, Kumar2019MelGAN, Prenger2019Waveglow, Yamamoto2020Parallel, Rohnke2020Parallel}.
However, none of these vocoders were thoroughly evaluated to claim universality.
In particular, it was not clear how well a speaker-independent vocoder performs on a target voice compared to a dedicated vocoder built specifically for that voice.
The closest setting to ours is the work of Lorenzo-Trueba et al. \cite{LorenzoTrueba2019Towards}, where a WaveRNN-based universal vocoder is capable of synthesising a wide range of speakers, styles, and conditions.
Unfortunately, Universal WaveRNN is autoregressive, and thus inherently slow in sample generation, posing significant difficulties for most real-time applications.
To the best of our knowledge, it remains unclear whether any non-autoregressive neural vocoders can be universal.

The contributions of this work are:
1) We present a universal neural vocoder based on Parallel WaveNet \cite{Oord2018Parallel}.
The key component of our universal vocoder is an additional conditioning network, called \textit{Audio Encoder}, which auto-encodes reference waveforms into utterance-level global conditioning.
2) Based on a large-scale evaluation, we show that the proposed universal vocoder significantly outperforms speaker-dependent vocoders overall.
It is capable of synthesising a wide range of in-domain and out-of-domain voices, speaking styles, and languages.
\OMIT{
	Results are consistent on both seen and unseen test scenarios.
}
3) We perform extensive benchmark studies on internal and open-source voices comparing several existing neural vocoder architectures in terms of naturalness and universality.
Results show that our universal vocoder has a clear advantage against other candidates.

\vspace{-0.4cm}%
\section{System description}
\vspace{-0.3cm}%
\label{sec:description}

\subsection{Parallel WaveNet}
\vspace{-0.2cm}%
\label{sec:pw}

Parallel WaveNet (PW) \cite{Oord2018Parallel} is a non-autoregressive neural vocoder architecture that transforms a sequence of input noise into audio waveforms in parallel.
It can synthesise samples very efficiently by fully exploiting the computational power of modern deep learning hardware.

In our early experiments, we found that PW trained on a multi-speaker dataset underperforms speaker-dependent PW. %(audio samples available on our demo page).
We conjecture that it can be empirically difficult to obtain a non-autoregressive vocoder that is able to faithfully reconstruct the phase structure of speech signals from speakers with diverse age, gender, speaking styles, and languages.

\vspace{-0.2cm}%
\subsection{Universal Parallel WaveNet with Audio Encoder}
\vspace{-0.2cm}%
\label{sec:uv}

Our universal vocoder is based on PW, that generates speech by conditioning on mel-spectrogram.
In order to make PW universal, we propose an additional conditioning network called \textit{Audio Encoder}, designed to explicitly model aspects of speech signals that are not provided by the mel-spectrogram conditioning.
The Audio Encoder encodes a reference waveform into a fixed-dimensional feature vector, which is then fed as utterance-level global conditioning into PW.
In the rest of the paper, we refer to PW with additional Audio Encoder conditioning as Universal Parallel WaveNet (UPW).

\begin{figure}[!tb]
	\centering
	\includegraphics[width=\columnwidth]{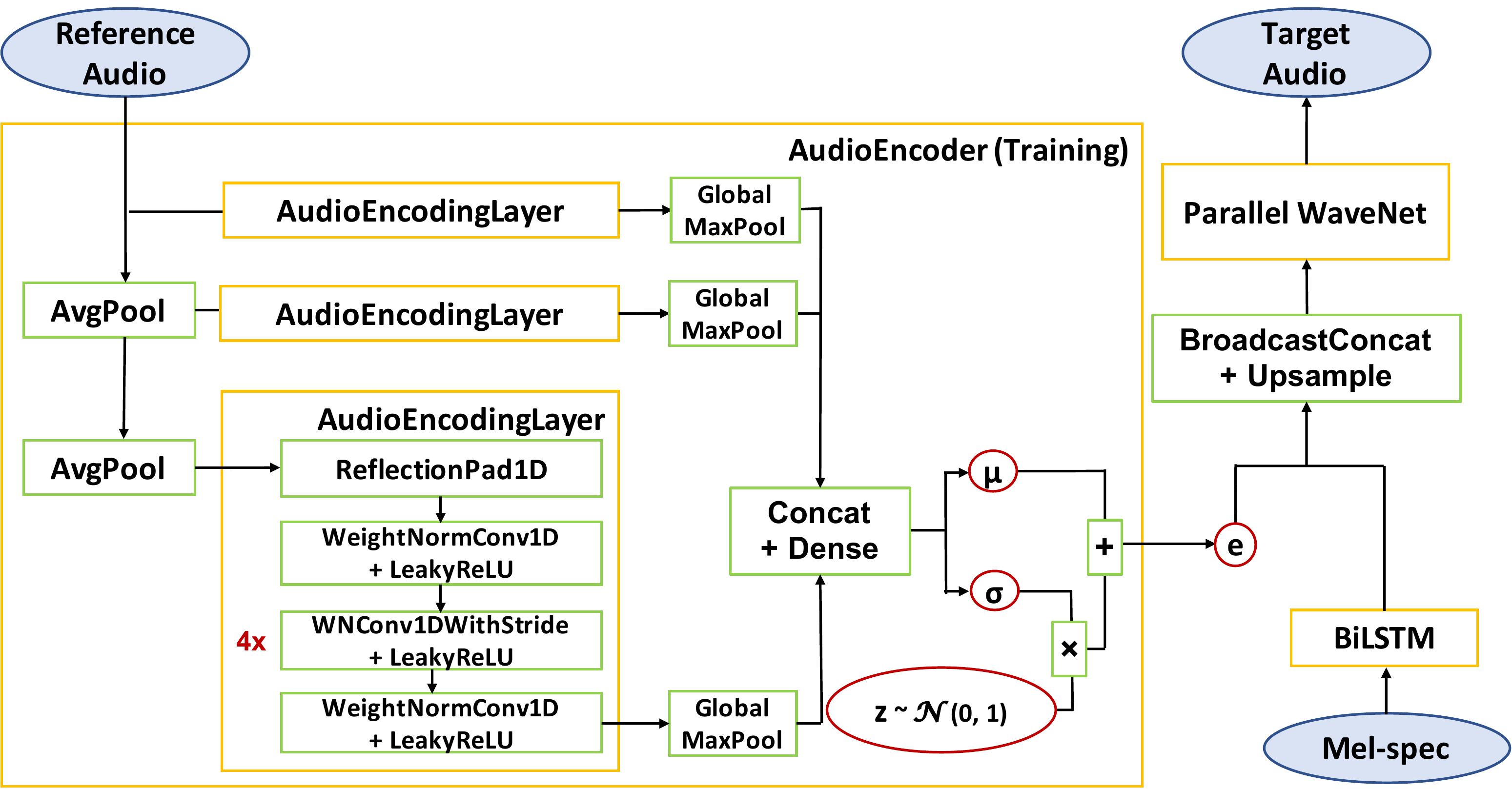}
	\caption{Universal Parallel WaveNet with Audio Encoder.}
	\label{fig:audioenc}
\end{figure}

The block diagram of the conditioning networks of the proposed UPW is shown in Figure \ref{fig:audioenc}.
The vocoder model has two conditioning networks --- an Audio Encoder and a mel-spectrogram conditioner.
First, the Audio Encoder is a multi-scale architecture of an audio feature extractor, heavily inspired by the design of MelGAN's discriminator \cite{Kumar2019MelGAN}.
It consists of 3 identical audio encoding layers that operate on different time-scales of the reference waveform, achieved by average pooling in between layers.
Each audio encoding layer uses a sequence of strided convolutional layers with a large kernel size \cite[Appendix A]{Kumar2019MelGAN}, where each convolutional layer is weight normalised and activated by Leaky ReLU, and each outputs 16 channels in the last layer followed by global max pooling and a dense layer.
To prevent information leakage to the vocoder, we applied amortised variational encoding \cite{Kingma2014Auto} to the output of Audio Encoder.
In the end, we obtain a total of 48-dimensional audio feature vector as an utterance-level global conditioning.
Second, we adopt the mel-spectrogram conditioner proposed by \cite{Arik2017Deep}, consisting of 2 bidirectional LSTMs with a hidden size of 128 channels.
The mel-spectrogram was extracted from ground-truth audio with 80 coefficients and frequencies ranging from 50 Hz to 12 kHz.
Finally, outputs from the two conditioning networks are broadcast--concatenated, and upsampled by repetition from frame-level (80 Hz) to sample-level (24 kHz).

During training, the target waveform is naturally used as the reference waveform.
This way, the entire architecture can be viewed as a conditioned Variational Auto-Encoder (VAE) on audio waveform, where the Audio Encoder network is an encoder and PW network is a decoder conditioned on mel-spectrogram.
Once the model is trained, doing inference with the proposed UPW requires a reference waveform input to the encoder, or use a pre-generated audio features in place of the output of Audio Encoder conditioner.
We investigated several inference strategies (e.g. using speaker-specific or style-specific centroid embedding in place of $\eb$), and found that using $\eb=\zerob$ generates high-quality audio (Figure \ref{fig:audioenc}).
In fact, $\eb=\zerob$ corresponds to the speaker-agnostic centroid embedding of the VAE prior distribution, and using it for inference is shown to improve generalisation of UPW to unseen voices.
This simple yet effective approach results in a UPW with little computational overhead compared to a basic PW, therefore both sharing the same real time factor in production.

Following the ``teacher-student'' training paradigm  \cite{Oord2018Parallel}, we first train a Universal WaveNet teacher, and then train a UPW student from it.
For the teacher network, we use 24 layers with 4 dilation doubling cycles, 128 residual/gating/skip channels, kernel size 3, and output distribution of a 10-component mixture of Logistics.
For the student network, we use [10, 10, 10, 30] flow layers with dilation reset every 10 layers, 64 residual channels, and no skip connections.
Both models were trained on sliced mel-spectrogram conditioning corresponding to short audio clips, with the Adam optimizer \cite{Kingma2015Adam} and a constant learning rate $10^{-4}$ until convergence.
The teacher uses batch size 64 and 0.3625s of audio clips.
The student uses batch size 16 and 0.85s of audio clips.
During distillation, student reuses the pre-trained conditioning networks of the teacher (including the Audio Encoder).
We found that training conditioning networks from scratch often leads to a worse student, a phenomenon also observed in \cite{Ping2019ClariNet}.

\OMIT{
	\OMIT{
		During distillation, we also incorporate the power loss as suggested by \cite{Oord2018Parallel}.
		Note that this means teacher will be training a conditioning network with Audio Encoder, and it will not be updated during distillation.
	}
	
	During training of the Audio Encoder, we applied amortised variational encoding \cite{Kingma2014Auto} to prevent too much information about the reference waveform leaking through the vocoder.
	This amounts to adding a loss term in the optimisation objective that penalises the KL divergence between a trainable Gaussian distribution of the latent audio features and a standard isotropic Gaussian prior.
	We also used monotonic annealing to improve the learning efficacy of variational bottlenecking \cite{Bowman2016Generating}.
	\OMIT{
		In fact, from the perspective of auto-encoding audio waveforms, the final architecture can be seen as a Conditional Variational Autoencoder \cite{Sohn2015Learning} of audio waveforms, where the conditioned decoder is a PW vocoder on mel-spectrogram.
	}
	The weight of the KLD loss is simply set equal to the weight of negative log-likelihood loss during teacher training.
}

\vspace{-0.4cm}%
\section{Experimental protocol}
\vspace{-0.3cm}%
\label{sec:protocol}

Our training and evaluation protocol was inspired by \cite{LorenzoTrueba2019Towards}, and adapted with a particular focus on universal vocoding of speech.
We collected a multi-speaker multi-lingual training set for the proposed universal vocoder.
It consists of 78 different internal, high-quality voices (20 males and 58 females) with approximately 3,000 utterances per speaker, and a total of 28 languages (including dialects) in 16 unique speaking styles (e.g. neutral, long-form reading, and several emotional styles in different degrees of intensity).
This training set was designed with the expectation that vocoders should be capable of synthesising a variety of voices, styles, and languages.
\OMIT{
	Audio signal was normalized to -23 LUFS, and dynamic range compression was used to reduce max true peak values to -7 dB.
	Crest factor of signal interquartile values concentrates between 6 and 7.
	Audio signal is encoded with 24 kHZ sampling rate and 16 bit floating point precision.
	Extracted mel-spectrograms were standardised channel-wise \TODO{should we cite what is standardisation?}.
}

\begin{table*}[!tb]
	\scriptsize
	\centering
	\begin{tabularx}{0.9\textwidth}{| c | c | Z | Z | Z | Z | Z | c |}
		\hline
		Section & Test sets & Recording quality & \# Voices (seen/unseen) & \# Styles (seen/unseen) & \# Languages (seen/unseen) & \# Utterances (all unseen) & Systems \\\hline
		\ref{sec:uv-vs-sd} & Internal & Very high & 24 (21/3) & 16 (12/4) & 13 (13/0) & 3124 & UPW, SDPW \cite{Oord2018Parallel} \\\hline
		\multirow{4}{*}{\ref{sec:uv-vs-others}} & Internal & Very high & 19 (15/4) & 2 (1/1) & 14 (14/0) & 1700 & \multirow{4}{*}{\shortstack[l]{UPW,\\UWRNN \cite{LorenzoTrueba2019Towards},\\PWGAN \cite{Yamamoto2020Parallel},\\WGlow  \cite{Prenger2019Waveglow}}} \\\cline{2-7}
		& LibriTTS-clean \cite{Zen2019LibriTTS} & High & 30 (0/30) & 1 (1/0) & 1 (1/0) & 300 & \\\cline{2-7}
		& LibriTTS-other \cite{Zen2019LibriTTS} & Medium & 30 (0/30) & 1 (1/0) & 1 (1/0) & 300 &   \\\cline{2-7}
		& Common Voice \cite{CommonVoiceDatabase} & Low & 300 (0/300) & 1 (1/0) & 15 (14/1) & 300 &  \\\hline
	\end{tabularx}
	\caption{Summary of test sets. Note that ``seen/unseen'' refers to whether they were exposed during training.}
	\label{tab:testsets}
\end{table*}

We perform analysis--synthesis on natural recordings, and design two types of evaluations on re-synthesised samples.\footnote{We also performed experiments on TTS samples using spectrograms generated by a Tacotron2-based architecture. Conclusions drawn from re-synthesised samples and TTS samples remain consistent.}
\vspace{-0.2cm}%
\begin{itemize}\setlength\itemsep{0em}
	\item In Section \ref{sec:uv-vs-sd}, we compare the proposed UPW with speaker-dependent PW (SDPW) on internal voices for which we have trained a high-quality SDPW.
	We show that the proposed vocoder is universal, in the sense that it does not show degradation when compared to speaker-dependent vocoders specific for each voice.
	\item In Section \ref{sec:uv-vs-others}, we benchmark UPW vs several other popular neural vocoder architectures in terms of universality.
	The competing vocoders include Universal WaveRNN (UWRNN) \cite{LorenzoTrueba2019Towards}, Parallel WaveGAN (PWGAN) \cite{Yamamoto2020Parallel}, and WaveGlow (WGlow) \cite{Prenger2019Waveglow}.
	All of these systems were re-trained on the same training set as our UPW, using an open-source implementation or reimplementing the default setup of each paper.
	We evaluate these vocoders on internal high-quality voices as well as external voices that are recorded in vastly different conditions.
\end{itemize}%
\vspace{-0.2cm}%
Table \ref{tab:testsets} summarises the statistics of the test sets in our experiments.
Note that test voices are selected such that they are balanced according to gender and age.

\OMIT{
	To ensure that the quality of a universal vocoder is not biased by the sheer volume of the training set, the single-speaker training set is each comparable in size to the multi-speaker training dataset.
	In fact, we found that naively increasing the size of the multi-speaker dataset set does not always lead to a better universal vocoder.
}

The naturalness perceptual evaluation was designed as a MUltiple Stimuli with Hidden Reference and Anchor (MUSHRA) \cite{Recommendation2001BS.}, where participants were presented with the systems being evaluated side-by-side, asked to rate them in terms of naturalness and audio quality (glitches, clicks, noise, etc.) from 0 (poorest) to 100 (best).
Each test utterance is evaluated by 10 to 15 listeners that are either native or educated speakers of the target language.
We also include recordings in all MUSHRA tests as the hidden upper-anchor system, and we do not force at least one 100 rated system.
\OMIT{
	Consequently, listeners may not always rate the recordings as the best-sounding system (i.e. relative MUSHRA scores of UPW can be higher than 100\%).
}

Paired two-sided Student T-tests with Holm-Bonferroni correction were used to validate the statistical significance of the differences between two systems at a $p$-value threshold of 0.05.
We refer to the ratio between the mean MUSHRA score of a system and natural recordings as relative MUSHRA (denoted by  Rel.)
Relative MUSHRA illustrates the gap between the system being evaluated with the reference.

\vspace{-0.4cm}%
\section{Results}
\vspace{-0.3cm}%
\label{sec:results}

\OMIT{
Do we want to show results on predicted mel-spectrograms? This is important from bussiness point of view, but not sure if matters much for community and will 2 times more space than going with only oracle mel spectrogram
}

\subsection{Comparison with speaker-dependent vocoders}
\vspace{-0.2cm}%
\label{sec:uv-vs-sd}

\begin{figure}[!tb]
	\centering
	\includegraphics[width=\columnwidth]{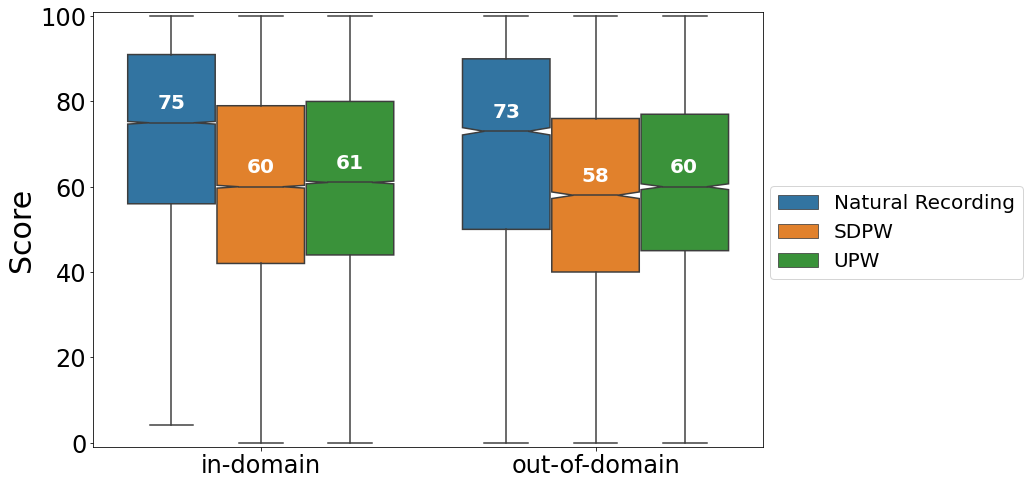}
	\caption{MUSHRA evaluation for comparison with speaker-dependent vocoders.}
	\label{fig:ntts-internal}
\end{figure}

\begin{table}[!tb]
	\scriptsize
	\centering
	\hspace*{-0.3cm}%
	\begin{tabular}{ c  c | c  c  c | c | c }
		\hline
		Voice (Sex) & Age & Rec. & SDPW & UPW & UPW Rel. & $p$-val. \\\hline
		\OMIT{abbey} British Eng. (F) & Adult & 71.64 &	65.69 &	\textbf{67.67} &	94.45\% &	\textbf{0.000} \\
		\OMIT{robin} Australian Eng. (M) & Adult & 73.52 &	\textbf{68.37} &	68.32 &	92.93\% &	1.000 \\
		\OMIT{alicia} Spanish (F) & Adult & 69.06 &	60.27 &	\textbf{61.17} &	88.58\% &	0.668 \\
		\OMIT{ira} Indian Eng. (F) & Adult & 77.19 &	62.22 &	\textbf{66.95} &	86.74\% & \textbf{0.000} \\
		\OMIT{colonel} *US Eng. (M) & Senior & 70.40 &	57.65 &	\textbf{60.12} &	85.40\% &	0.201 \\
		\OMIT{justin} *US Eng. (M) & Child & 62.31 &	51.26 &	\textbf{51.99} &	83.43\% &	1.000 \\
		\OMIT{matthew} US Eng. (M) & Adult & 68.58 &	52.63 &	\textbf{55.46} &	80.87\% &	0.105 \\
		\OMIT{chloe} French (F) & Senior & 72.53 &	54.82 &	\textbf{56.35} &	77.69\% &	\textbf{0.002} \\
		\OMIT{candela} US Spanish (F) & Adult & 73.71 &	48.07 &	\textbf{48.37} &	65.62\% &	1.000 \\\hline\hline
		\multicolumn{2}{c|}{Overall} & 69.68 &	57.92 &	\textbf{58.70} &	84.24\% &	\textbf{0.000} \\\hline
	\end{tabular}
	\caption{MUSHRA scores on internal voices (unseen marked by *). $p$-value signifies the difference between UPW vs SDPW. Note that the voices are selected so that the relative MUSHRA of UPW is evenly distributed within the range from highest to lowest.}
	\label{tab:ntts-voices}
\end{table}

\begin{table}[!tb]
	\scriptsize
	\centering
	\begin{tabular}{ c | c  c  c | c | c }
		\hline
		Style & Rec. & SDPW & UPW & UPW Rel. & $p$-val. \\\hline
		Emotional & 71.59 &	60.74 &	\textbf{61.40} &	85.76\% & 0.462 \\
		Neutral & 69.13 &	58.53 &	\textbf{58.73} &	84.95\% & 0.500 \\
		Conversational & 58.65 &	43.54 &	\textbf{47.61} &	81.18\% &	\textbf{0.002} \\
		Long-form reading & 68.60 &	\textbf{56.69} &	55.46 &	80.85\% & 0.814 \\
		News briefing & 75.24 &	56.29 &	\textbf{59.86} &	79.55\% & \textbf{0.000} \\
		Singing & 71.94 &	49.96 &	\textbf{56.87} &	79.06\% & \textbf{0.000} \\
		\hline
	\end{tabular}
	\caption{MUSHRA scores on typical styles. $p$-value signifies the difference between SDPW vs UPW.}
	\label{tab:ntts-styles}
\end{table}

In this section, we compare the proposed universal vocoder UPW with speaker-dependent vocoder (SDPW).
MUSHRA evaluations are carried out on 24 high-quality internal voices in distinct speaking styles and languages, using Amazon's internal evaluation platform where listeners are professionally trained for voice quality assessment.

Results show that UPW significantly outperforms SDPW overall with a relative MUSHRA of 84.24\% vs 83.12\% ($p$-value $=0$).
As we break down to each voice, UPW shows statistically significant improvement to SDPW on 7 voices, and is comparable to SDPW on 17 voices without statistically significant difference.
Table \ref{tab:ntts-voices} lists some voices for which the relative MUSHRA achieved by UPW is evenly distributed within the range from highest (94.45\%) to lowest (65.62\%).
This is a very strong result for the proposed universal vocoder, since not only can it avoid degradation compared to speaker-specific vocoders on all tested voices, but also it improves the vocoding quality on many voices by using information contained in the speech signal of related speakers.
Moreover, Figure \ref{fig:ntts-internal} shows that UPW consistently outperforms SDPW on voices and speaking styles that were either exposed during training (in-domain) or not (out-of-domain).

We now focus our evaluation on different speaking styles, as we found that it is usually challenging to vocode highly expressive speech even for a well-trained SDPW.
Table \ref{tab:ntts-styles} summarises some typical styles in our evaluation.
We find that UPW is comparable to SDPW on neutral, emotional (e.g. excited, disappointed), and long-form reading  style.
For some expressive styles such as conversational, news briefing, and singing style, UPW statistically significantly outperforms SDPW.
In particular, the most challenging style we consider in our evaluation is singing, because we believe it is the most expressive type of speech.
While both UPW and SDPW indeed achieved the least relative MUSHRA on this style (79.06\% vs 69.46\%), UPW sees the greatest improvement from SDPW.
Notably, UPW outperforms SDPW on singing style by 6.91 MUSHRA points on average, closing the gap between recordings and SDPW by 31.44\%.
This strongly evidences the superiority of the proposed universal vocoder.

\vspace{-0.2cm}%
\subsection{Comparison with other multi-speaker vocoders}
\vspace{-0.2cm}%
\label{sec:uv-vs-others}

In this section, we compare UPW with other popular neural vocoder architectures, including Universal WaveRNN (UWRNN) \cite{LorenzoTrueba2019Towards}, Parallel WaveGAN (PWGAN) \cite{Yamamoto2020Parallel} and WaveGlow (WGlow) \cite{Prenger2019Waveglow}.
Note that MUSHRA evaluations in this section are carried out by Clickworker \cite{Clickworker}.

\vspace{-0.3cm}%
\paragraph*{Internal voices.}

\OMIT{
	\begin{figure}[!tb]
		\centering
		\includegraphics[width=\columnwidth]{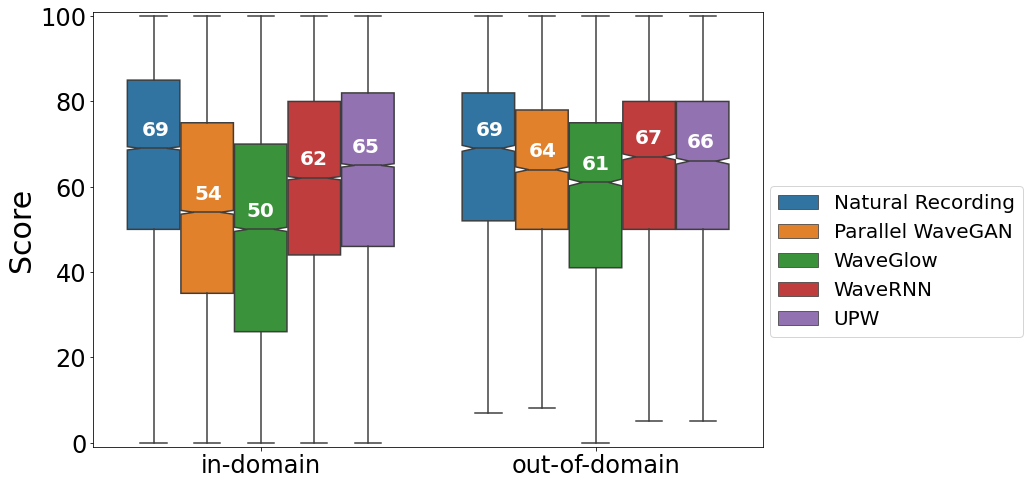}
		\caption{MUSHRA evaluation for comparison with other multi-speaker vocoders.}
		\label{fig:nonntts-internal}
	\end{figure}
}

\begin{table}[!tb]
	\scriptsize
	\centering
	\hspace*{-0.2cm}%
	\begin{tabular}{ c | c  c  c  c  c | c }
		\hline
		Voice (Sex) & Rec. & PWGAN & WGlow & UWRNN  & UPW & UPW Rel. \\\hline
		\OMIT{georgio} Italian (M) & 65.97 &	55.05 &	50.67 &	59.46 &	\textbf{65.83} &	99.78\% \\
		\OMIT{gwynethcy} *Welsh (F) & 65.67 &	57.12 &	57.25 &	63.50 &	\textbf{65.17} &	99.23\% \\
		\OMIT{seoyeon} Korean (F) & 70.60 &	64.90 &	56.20 &	60.26 &	\textbf{68.16} &	96.54\% \\
		\OMIT{ewa} Polish (F) & 67.21 &	57.83 &	47.39 &	64.35 &	\textbf{64.68} &	96.24\% \\
		\OMIT{mathieu} French (M) & 65.14 &	51.34 &	47.68 &	57.94 &	\textbf{61.20} &	93.96\% \\
		\OMIT{hans} German (M) & 68.61 &	60.86 &	50.93 &	\textbf{66.27} &	61.57 &	89.75\% \\
		\OMIT{jacek} Polish (M) & 65.12 &	54.61 &	46.23 &	\textbf{58.69} &	53.28 &	81.82\% \\\hline\hline
		Overall & 66.81 &	56.02 &	50.09 &	61.83 &	\textbf{63.35} &	94.82\% \\\hline
	\end{tabular}
	\caption{MUSHRA scores on internal voices (unseen marked by *). All speakers are adult. Note that the voices are selected so that the relative MUSHRA of UPW is evenly distributed within the range from highest to lowest.}
	\label{tab:nonntts-voices}
\end{table}

\OMIT{
	\begin{table}[!tb]
		\scriptsize
		\centering
		\begin{tabular}{ c | c  c  c  c  c | c }
			\hline
			Style & Rec. & PWGAN & WGlow & UWRNN & UPW & UPW Rel. \\\hline
			*Character & 66.77 &	65.41 &	64.05 &	65.81 &	\textbf{67.36} &	100.90\% \\
			Neutral & 66.81 &	55.4 &	49.17 &	61.57 &	\textbf{63.08} &	94.42\% \\
			\hline
		\end{tabular}
		\caption{MUSHRA scores on typical styles (unseen marked by *).}
		\label{tab:nonntts-styles}
	\end{table}
}

We first evaluate competing systems on 19 high-quality internal voices.
The results in Table \ref{tab:nonntts-voices} show that UPW is the best-performing vocoder overall ($p$-value $=0$), achieving the highest average relative MUSHRA of 94.82\% among all four competing vocoders.
Some voices are listed for which the relative MUSHRA achieved by UPW is evenly distributed within the range from highest (99.78\%) to lowest (81.82\%).
Compared to the other non-autoregressive candidates, UPW statistically significantly outperforms WGlow on all 19 tested voices.
UPW statistically significantly outperforms PWGAN on 16 voices, and both systems are comparable on 3 voices without statistically significant difference.
Compared to the autoregressive candidate, UPW is statistically significantly better than UWRNN on 13 voices, both are comparable on 2 voices, but UPW underperforms UWRNN on 4 voices.
However, it is worth noting that, due to its autoregressive nature, UWRNN has an inference speed that is slower than UPW typically by orders of magnitude.

%\vspace{-0.3cm}%
\paragraph*{External voices.}

\OMIT{
	\begin{figure*}[!htb]
		\small
		\centering
		\begin{minipage}[t]{0.3\linewidth}
			\centering
			\centerline{\includegraphics[width=\textwidth]{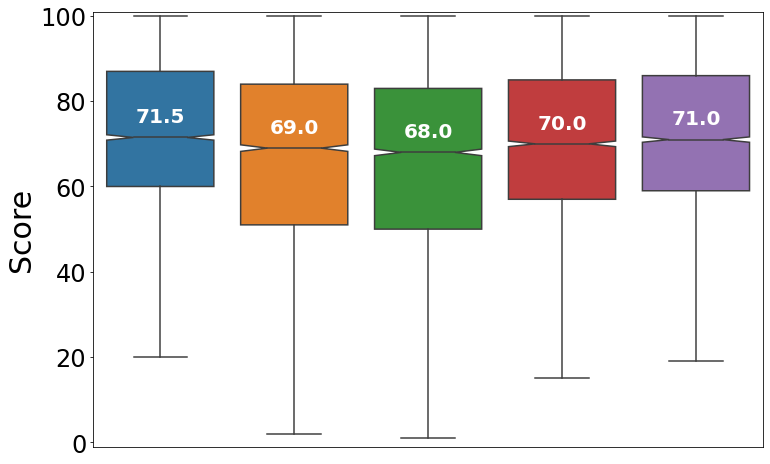}}
			\centerline{(a) LibriTTS-clean.}%\medskip
		\end{minipage}%
		\begin{minipage}[t]{0.3\linewidth}
			\centering
			\centerline{\includegraphics[width=\textwidth]{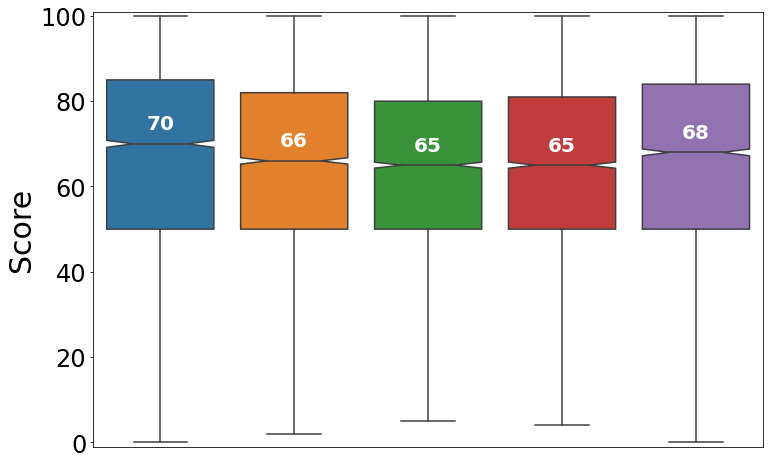}}
			\centerline{(b) LibriTTS-other.}%\medskip
		\end{minipage}%
		\begin{minipage}[t]{0.4\linewidth}
			\centering
			\centerline{\includegraphics[width=\textwidth]{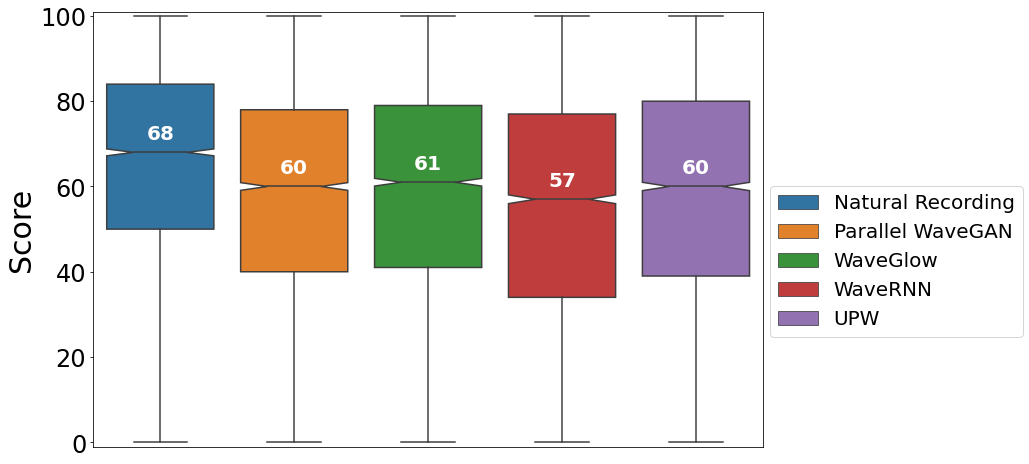}}
			\centerline{(c) Common Voice.}%\medskip
		\end{minipage}%
		\caption{MUSHRA evaluation for robustness to external voices.}
		\label{fig:external}
	\end{figure*}
}

\begin{table}[!tb]
	\scriptsize
	\centering
	\hspace*{-0.3cm}%
	\begin{tabular}{ c | c  c  c  c  c | c }
		\hline
		Dataset & Rec. & PWGAN & WGlow & UWRNN  & UPW & UPW Rel. \\\hline
		LibriTTS-clean & 70.42 &	67.40 &	66.72 &	68.30 &	\textbf{69.56} &	98.77\% \\
		LibriTTS-other & 68.91 &	65.04 &	64.15 &	63.83 &	\textbf{67.28} &	97.64\% \\
		Common Voice & 64.84 &	57.84 &	\textbf{58.67} &	54.87 &	58.07 &	89.56\% \\
		\hline
	\end{tabular}
	\caption{MUSHRA scores on external voices.}
	\label{tab:external}
\end{table}

We further study the robustness of the pre-trained UPW on open-source voices.
To this end, we prepared three test sets with decreasing recording quality: LibriTTS-clean \cite{Zen2019LibriTTS}, LibriTTS-other \cite{Zen2019LibriTTS}, and Common Voice \cite{CommonVoiceDatabase}.
Note that LibriTTS is a multi-speaker corpus of English speech in audiobook reading style, and Common Voice is a database of multi-lingual multi-speaker user recording.
The results in Table \ref{tab:external} show that UPW is a top-performing system across all three sets of external voices.
(a) On high-quality voices (LibriTTS-clean), UPW consistently outperforms all other systems, achieving a relative MUSHRA of 98.77\%. This implies that UPW can generalise well to out-of-domain voices when the recording conditions are studio-quality.
(b) On medium-quality voices (LibriTTS-other), UPW has a clear advantage over other systems. This strongly suggests that UPW is still capable of synthesising naturally sounding speech in the presence of a reasonable level of noise.
(c) On low-quality voices (Common Voice), WGlow and UPW are in fact comparably good ($p$-value $=0.15$), while UWRNN is the least robust to recording conditions where a significant amount of background noise is present in low-quality speech.

\vspace{-0.3cm}%
\section{Conclusion}
\vspace{-0.3cm}%
\label{sec:conclusion}

In this work, we presented a universal neural vocoder based on Parallel WaveNet, trained on a multi-speaker multi-lingual speech dataset.
It is capable of synthesising a wide range of voices, styles, and languages, and particularly suitable for scaling up production of real-time TTS.
The key component of the proposed universal vocoder is an additional conditioning network called Audio Encoder, which auto-encodes reference waveforms into utterance-level global conditioning.
Based on large-scale evaluation, our universal vocoder outperforms speaker-dependent vocoders overall.
We have also reported extensive studies benchmarking several existing neural vocoder architectures in terms of naturalness and universality, and showed that our universal vocoder has a clear advantage of being non-autoregressive and superior in terms of quality in a vast majority of cases.

There are still interesting research directions we will leave to future work.
First, it is interesting to generalise the proposed Audio Encoder to encode reference waveforms into local conditioning, which can better represent the localised features of speech signals.
Second, we can study whether the proposed Audio Encoder would also benefit multi-speaker training of other neural vocoders such as WaveGlow \cite{Prenger2019Waveglow} and Parallel WaveGAN \cite{Yamamoto2020Parallel}.
Third, it is worth investigating how our universal vocoder performs on challenging vocoding scenarios, such as overlapping voices, and non-speech vocalisations (e.g. shouts, breath, reverberation).
\OMIT{
	For example, this would require to compose a training set that can cover non-speech vocoding without degrading the quality on speech vocoding.
}

%\vspace{-0.3cm}%
%\paragraph*{Acknowledgement.}
%
%We thank Alexis Moinet and Vatsal Aggarwal for providing insightful research discussions.
%

\vfill\pagebreak

\bibliographystyle{IEEEbib}
\bibliography{refs}

\end{document}